# Direct ultrafast parametric amplification pumped by a picosecond thin-disk laser

Jing Wang[1], Peng Yuan[1], Dongfang Zhang[1], Guoqiang Xie[1], Kainan Xiong[2], Xiaoniu Tu[2], Yanqing Zheng[2,3], Jingui Ma[1], and Liejia Qian[1,4]

[1] Key Laboratory for Laser Plasmas (MOE), Collaborative Innovation Center of IFSA (CICIFSA), School of Physics and Astronomy, Shanghai Jiao Tong University, Shanghai 200240, China

[2] Shanghai Institute of Ceramics, Chinese Academy of Sciences, Shanghai 201800, China

[3] School of Material Science and Chemical Engineering, Ningbo University, Ningbo, Zhejiang 315211, China.

[4] Tsung-Dao Lee Institute, Shanghai Jiao Tong University, Shanghai 200240, China

Optical quadratic nonlinearity is ultrafast in nature, while parametric interaction usually manifests only the broadband characteristic [1]. Enormous progress has been made toward broadband phase-matching for parametric amplification and wide applications [1-3]. In existing devices of broadband parametric amplification, the power efficiency of conversion is restricted to approximately the energy efficiency, and the desire for the signal power enhancement necessitates additional pulse compression after amplification [4-10]. Here, we demonstrate ultrafast parametric amplification having an extraordinary power efficiency of 1155%; this allows the generation of intense femtosecond pulses without the need for a pulse stretcher and compressor. Direct femtosecond signal amplification by picosecond pumping is enabled by an ultrafast parametric environment in which the pre-delayed signal of faster speed gradually overtakes and effectively depletes the pump of slower speed as they propagate in a nonlinear crystal. The demonstrated technique should lead to breakthroughs in ultrafast lasers as well as applications.

Signal parametric amplification through three-wave mixing is a quadratic nonlinear optical process under a strong pump [1]. It takes a variety of forms from narrowband to broadband, which can be characterized by the phase-matching dispersion property. In any parametric amplification device, the zeroth order term of phase matching, i.e., phase-velocity matching, must be fulfilled for efficient conversion. For broadband parametric amplification, the first-order phase-matching dispersion, i.e., the group-velocity mismatch ($GVM_{si}$) between the signal and idler, must be further compensated [2, 3]. The concept of broadband parametric amplification is well suited to femtosecond pump pulses. Limited by the available power of femtosecond Ti:sapphire lasers at a kHz repetition rate, such femtosecond-pumped parametric amplification typically generates tunable femtosecond pulses with an energy of approximately 100 µJ or less [4–6].

The main issue for high-power femtosecond lasers with parametric amplification is related to the temporal matching between the pulses of the femtosecond signal and high-energy pump. Unlike conventional gain media (e.g., Ti:sapphire) that can accumulate and store energy, signal parametric amplification is an instantaneous nonlinear process and directly acquires the energy from the pump. This means that the energy has to be stored in the pump laser system, and the pump laser needs to be at least a picosecond in duration to generate energetic pulses from well-established gain media such as Yb-doped laser crystals [11, 12]. With a long pump pulse, to achieve efficient energy extraction, stretching the femtosecond signal pulse before parametric amplification is a traditional approach. This scheme was termed optical parametric chirped-pulse amplification (OPCPA) [7, 8]. Today, OPCPA is at the center of ultrafast lasers, with utmost importance for time-resolved optical spectroscopy and high-field physics [9, 10, 13, 14]. However, OPCPA still has problems. As they typically require angular dispersion to constitute a pulse stretcher and compressor [15], conventional OPCPA systems tend to be relatively complex and large devices that are difficult to align and miniaturize [16]. More seriously, the involved diffraction gratings and spatiotemporal couplings arising from angular dispersion may degrade the crucial laser parameters of carrier-envelope phase stability, focused intensity and pulse contrast [17–20].

Here, we introduce an ultrafast parametric environment suitable for directly amplifying a femtosecond signal. As illustrated in Fig. 1a, such direct ultrafast parametric amplification (DUPA) consists of a signal pulse with femtosecond duration $\tau_s$, a pump pulse with picosecond duration $\tau_p$, and a nonlinear crystal with thickness L and nonvanishing $GVM_{ps}$ between the pump and signal. When the ultrafast parametric environment satisfies a temporal condition of $L \times GVM_{ps} \approx \tau_p$, the whole pump pulse contributes to the femtosecond signal amplification. For the situation of wide interest of a green-light pump and a near-infrared signal, most nonlinear crystals have an appropriate $GVM_{ps} \approx 1$ ps cm$^{-1}$, which accommodates a range of picosecond pump pulses from 0.5 ps to 5 ps.

DUPA can be made efficient without the need for a pulse stretcher and compressor. Under the typical condition, the numerical simulation of DUPA shows a high energy efficiency of 39% from pump to signal (Fig. 1b; see Methods for details). The intriguing dynamics of DUPA can be better illustrated by the residual pump pulse. Since the instantaneous interaction only occurs in a thin slice of $L_s \approx \tau_s / GVM_{ps}$ inside the crystal or in a femtosecond slice of $\tau_s$ within the pump pulse, the depletion of the pump pulse begins with a characteristic temporal dip of approximately $\tau_s$ and gradually evolves into the major leading portion (the insets in Fig. 1b). In other words, DUPA can be equivalent to a series of femtosecond-pumped parametric amplification units in the picosecond window of the pump pulse.

A nonlinear crystal with proper $GVM_{ps}$ and nonlinear coefficient $d_{eff}$ is the key material required for implementation of DUPA. In this first experimental demonstration, we take advantage of a $Sm^{3+}$-doped yttrium calcium oxyborate (Sm:YCOB) crystal (see Methods). The Sm:YCOB crystal has attractive properties of $GVM_{ps} = 1$ ps cm$^{-1}$ between the pump (515 nm) and signal (778 nm), nonlinear coefficient $d_{eff} = 1.39$ pm V$^{-1}$, and idler (1523 nm) absorption $\alpha \approx 3$ cm$^{-1}$ for enhancement of the signal efficiency [21–23].

The experimental setup of DUPA was simple, in which a 3.9-cm-thick Sm:YCOB crystal was inserted into the beam intersection area of the pump and signal lasers (see Methods). To achieve $GVM_{si} = 0$ and hence a maximized bandwidth, a noncollinear phase-matching geometry was adopted by setting a 65 mrad intersection angle between the pump and signal. Figure 2a gives the measured gain spectrum at a fixed crystal orientation. To optimize the small-signal gain, the proper time delay between the signal and pump was optimized such that the femtosecond signal and the pump pulse peak overlapped in the crystal middle, i.e., symmetric mapping of the time delay to the crystal position. With a pump intensity of 6.9 GW cm$^{-2}$, DUPA had a high gain of 3×10$^7$ in a spectral width of 32 nm. As the signal seed energy increased, DUPA fell into the saturated amplification regime. A signal energy efficiency of 30% was obtained under a fixed pump energy of 3.40 mJ and the strongest available seed energy of 10.5 μJ (Fig. 2b). Meanwhile, DUPA produced the highest signal energy of 1.02 mJ by amplification. Even higher efficiency close to the theoretical limit should be possible by further increasing the signal seed. In the saturated amplification, the symmetric mapping between the time delay and crystal position remained unchanged, which rendered the optimal efficiency. If the temporal condition of ultrafast parametric environment were not satisfied or equivalently only part of the pump pulse could interact with the femtosecond signal pulse within the crystal, then high-efficiency DUPA would not be possible, and the required time delay would cause the symmetric mapping to slightly deviate toward the crystal exit.

DUPA directly delivered the femtosecond amplified signal output at the crystal exit. In the experimental condition with a signal bandwidth of 30 nm, the crystal dispersion of approximately 6300 fs$^2$ needs to be precompensated by a Dazzler acousto-optic pulse shaper. Consequently, the amplified signal output at the crystal exit was a nearly transform-limited pulse with a 65 fs duration and a 30 nm bandwidth, similar to the values without amplification (Fig. 3a and Fig. 3b). Considering the 30% energy efficiency and 38.5-fold reduction in the pulse duration from the pump to the signal, we deduced a power efficiency of 1155% from the pump to the signal. Notably, the attainable reduction factor of the pulse duration linearly increases with GVM$_{ps}$, and hence, the amplified femtosecond signal can be much stronger than the pump pulse if a crystal with large GVM$_{ps}$ is available. We further verified the faithful high efficiency via pump pulse depletion. Both the incident and depleted pump pulses were measured (Fig. 3c and Fig. 3d), which clearly indicated that DUPA principally depleted the leading portion of the pump pulse. Meanwhile, the blue portion of the pump spectrum was accordingly depleted. This can be attributed to the imposed negative chirp that stretched a transform-limited 0.35 ps pulse to a 2.5 ps pump pulse.

DUPA is not only broadband but also ultrafast. The unusual dynamics of DUPA results from the interplay of the femtosecond signal pulse duration, pump pulse chirp and nonvanishing $GVM_{ps}$. In the experimental study, the residual pump pulse and spectrum at each position inside the crystal could not be measured. Here, we introduce an alternative method. The dynamic evolution of the pump spectrum upon depletion was observed at the crystal exit by scanning the time delay between the signal and pump, as shown in Fig. 4a. Specifically, a characteristic dip in the first frame of the residual pump spectrum is the signature of the beginning of depletion, which implies that pump depletion only occurs at (or very close to) the crystal exit. With increasing time delay relative to the first frame, the femtosecond signal will meet the pump pulse peak at an earlier position inside the crystal, and hence, the signal amplification will deplete more spectral components of the pump. The characteristic spectral shift ($\Delta\lambda$) toward the blue edge has a linear dependence on the increase in the time delay ($\Delta_T$), in which the slope of $\Delta_T / \Delta\lambda \approx -2.2$ ps nm$^{-1}$ defines the pump pulse chirp.

$GVM_{ps}$ is a crucial parameter that governs the performance of DUPA. Reciprocally, the value of $GVM_{ps}$ can be deduced from the full evolution of the pump depletion. First, the seed signal energy was purposely decreased to let the characteristic dip in the first frame occur in the center of the pump spectrum (Fig. 4a and Fig. 4b). Under such a specific condition, the femtosecond signal pulse coincided with the pump pulse peak at the crystal exit. Thus, the absolute time interval $T_0$ between the signal and pump before the crystal entrance simply represents the overall group-velocity mismatch L × $GVM_{ps}$ (Fig. 4c). Second, the time interval $T_0$ can be deduced from the pump spectrum evolution over a large range of time delays. As shown in Fig. 4b, the spectral depletion always remained maximized from $\Delta_T = 0.8$ ps to $\Delta_T = 1.8$ ps and then gradually decreased until the recovery of the initial pump spectrum. The time interval $T_f \approx 2$ ps can be reasonably judged for the specific condition with $\Delta_T = 1.8$ ps. In accordance with the timing diagrams in Fig. 4c, the combination of the above two experimental steps yields $T_0 \approx 3.8$ ps and $GVM_{ps} \approx 1$ ps cm$^{-1}$. Previously, $GVM_{ps}$ was often estimated from the refractive index of crystals, which may deviate from the real value, particularly for a new kind of crystal without accurate refractive index data. The demonstrated method for experimentally characterizing $GVM_{ps}$ is helpful to the design of DUPA.

For efficient energy extraction from high-energy-storage materials by femtosecond pulses, application of the chirped-pulse amplification (CPA) technique has been widely accepted to be required to mitigate the third-order nonlinear effects associated with high intensity. Contrary to this belief, here, the femtosecond signal was directly amplified by DUPA. The experimental system was highly reliable, in which no nonlinear breakdown occurred even under a high repetition rate of 1 kHz. This is surprising but can be understood from the high gain property of DUPA that allows a short interaction length. In the experiment, the average gain coefficient was as high as 4.4 cm$^{-1}$, which is approximately 100 times larger than those of Ti:sapphire and Nd:glass CPA lasers [24]. Consequently, the third-order nonlinear effects can be relatively mitigated by 100 times, and hence, the CPA technique may no longer be needed in DUPA. In fact, the estimated nonlinear phase shifts in the DUPA crystal with and without amplification were both below the critical value of π for nonlinear breakdown [25].

In summary, we have demonstrated picosecond-pumped DUPA without a pulse stretcher and compressor. A high energy efficiency of 30% has been achieved by the ultrafast parametric environment of an idler absorption nonlinear crystal and velocity-mediated temporal matching between the femtosecond signal and picosecond pump. The DUPA scheme is well suited to picosecond pump lasers, which can be regarded as another new class of ultrafast laser devices since the advent of femtosecond-pumped parametric amplification and OPCPA. A remarkable application scenario is integration with the latest thin-disk technology. Notably, DUPA has considerable potential for high performance. Initial DUPA results obtained in the proof-of-principle experiment, i.e., 1 mJ and 65 fs pulses at a 1 kHz repetition rate, already rival the best achieved with commercial femtosecond-pumped parametric amplifier devices (e.g., OPerA SOLO, Coherent). With higher pump energy from state-of-the-art thin-disk lasers [11, 12], we expect that, in principle, scaling of the pulse energy by more than two orders of magnitude will be possible. Through its ability to produce terawatt femtosecond pulses at a 1 kHz repetition rate over a very wide spectral range, DUPA holds promise for surpassing the performance of current femtosecond Ti:sapphire systems

[26]. Thus, DFPA in combination with thin-disk technology should lead to breakthroughs in ultrafast lasers as well as applications.

## Methods

Numerical methods and parameters. Standard nonlinear coupled-wave equations in full dimensions were adopted for the DUPA simulation under both phase matching ($\Delta k = 0$) and group-velocity matching ($GVM_{si} = 0$), where idler absorption coefficient $\alpha = 3$ cm$^{-1}$ and $GVM_{ps} = 1$ ps × cm$^{-1}$ were included and other linear effects were neglected. The wavelengths of the three interacting waves were set to 515 nm (pump), 778 nm (signal) and 1523 nm (idler). The simulation assumed a Gaussian pump in both time and space, $I_p(x,y,t) = I_{p0}\exp[-(2\sqrt{\ln 2}\ x/\sigma_p)^2]\exp[-(2\sqrt{\ln 2}\ y/\sigma_p)^2]\exp[-(2\sqrt{\ln 2}\ t/\tau_p)^2]$, where $I_{p0}$ is the peak intensity and $\sigma_p = 0.5$ cm ($\tau_p = 2.5$ ps) is the beam width (pulse duration). The nonlinear length $L_{nl}$ was fixed at 0.22 cm, which determines the product of pump field $E_{p0}$ and nonlinear coefficient $d_{eff}$ [10]. In addition, the Gaussian signal seed had the specifications of beam radius $\sigma_s = \sigma_p$, pulse duration $\tau_s = 0.05\ \tau_p$ and fixed intensity $I_{s0} = 0.05\ I_{p0}$.

Sm:YCOB crystal. The 30%-Sm$^{3+}$-doped YCOB crystal was grown in an induction-heating Czochralski furnace in a gaseous atmosphere of $N_2$ and 1 vol% $O_2$ using a crystallographic b-oriented YCOB bar with dimensions of 0.4 × 0.4 × 4 cm$^3$ as the seed. The specially grown Sm:YCOB crystal had an average absorption of $\alpha = 3$ cm$^{-1}$ around the wavelength of 1523 nm and was cut in the type-I phase-matching orientation (θ=114.1°, φ=39.8°) to obtain the largest nonlinear coefficient. In the DFPA experiment, an antireflection-coated Sm:YCOB crystal with a 3.9 cm thickness and a 1 × 1 cm$^2$ surface was applied.

Pump and signal lasers. The 515 nm pump laser had a 2.5 ps pulse duration, a 0.5 cm beam width, a 3.40 mJ energy and a 6.9 GW cm$^{-2}$ intensity, which was obtained by imposing negative dispersion and second-harmonic generation on a picosecond thin-disk Yb:YAG laser at a 1 kHz repetition rate (Vary Disk VE150, Dausinger-Giesen GmbH). The femtosecond laser source consisted of a Kerr-lens mode-locked Ti:sapphire oscillator (Venteon CEP5, Laser Quantum), a Dazzler acousto-optic pulse shaper (HR800, Fastlite), and three OPCPA stages at a 1 kHz repetition rate. The output basically covered an ultrabroad spectrum from 670 to 970 nm. With the pulse shaper activated, the required 778 nm femtosecond signal with a 30 nm bandwidth was spectrally selected, amplified to 10.5 µJ and temporally compressed to 60 fs at the exit of the Sm:YCOB crystal.

Experimental details. The experimental setup of DUPA was as simple as the well-known second-harmonic generation device, in which a 3.9-cm-thick Sm:YCOB crystal was pumped with a 65 mrad noncollinear angle to the signal. To measure the gain spectrum under the fixed phase-matching condition for the 778 nm signal, the signal seed had a purposely ultralow energy of 0.1 pJ, a bandwidth of 30 nm and a tunable wavelength range from 745 nm to 815 nm. In the measurement of the signal energy efficiency, the signal seed had a fixed wavelength of 778 nm, a bandwidth of 30 nm and a varied energy up to 10.5 µJ. In addition, femtosecond pulses were measured by a frequency-resolved optical gating (FROG) device (GRENOUILLE 8-20-USB, Swamp Optics), while picosecond pulses were measured by their cross-correlation with the femtosecond signal. All pulse spectra were measured by a double-dispersion monochromator spectrograph (SPMSDD1004, SOL instruments).

## Acknowledgements

This study was partially supported by grants from the Natural Science Foundation of China (91850203, 61905142, 61975120, 61727820, 51832009).

## Author contributions

L.Q. and J.W. contributed the concept of direct ultrafast parametric amplification to the team. J.W. and P.Y. carried out the simulation and experiment. Y.Z., K.X. and X.T. designed and grew the Sm:YCOB crystal. J.M., D.Z. and G.X. oversaw the experimental area and contributed to the design. All of the authors contributed to the preparation of the manuscript.

## Additional information

Competing financial interests: The authors declare no competing financial interests.

Figure Legends

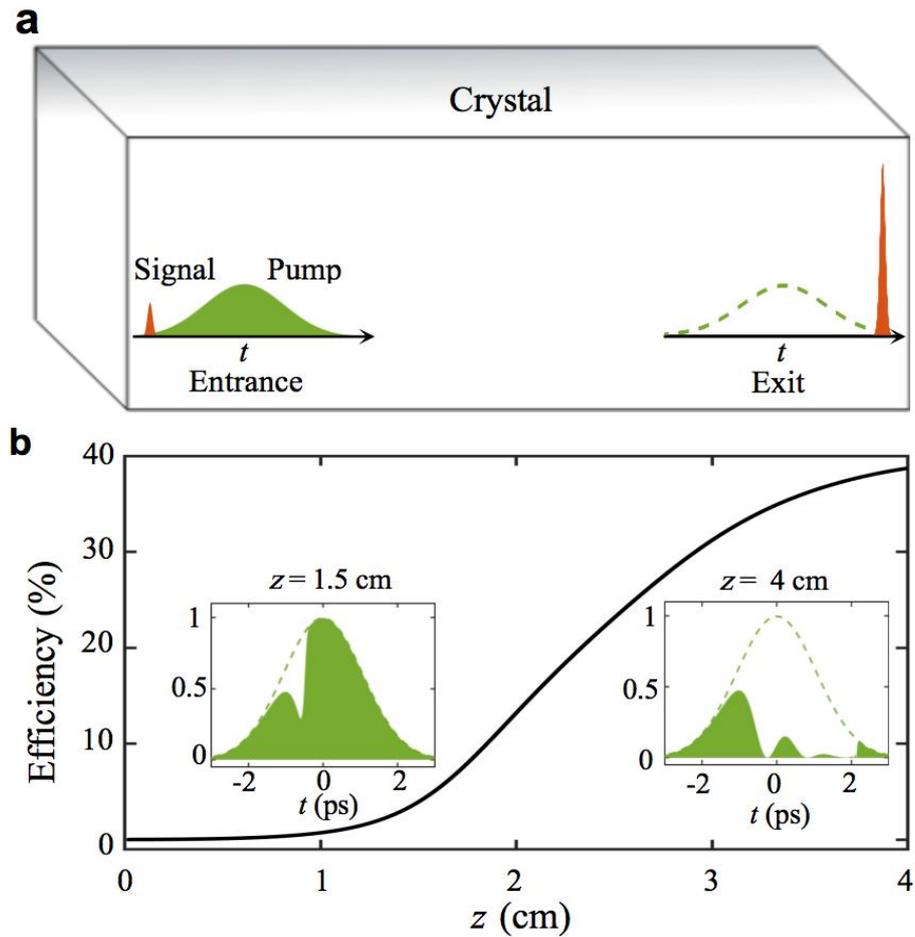

Figure 1 | Principle of DUPA. a, Schematic diagram of the ultrafast parametric environment. The pre-delayed femtosecond signal of faster speed gradually overtakes and effectively depletes the picosecond pump of slower speed as they propagate in the nonlinear crystal. b, Calculated efficiency from the pump to the signal. Insets: two residual pump pulses in the beam center at z = 1.5 cm and 4 cm.

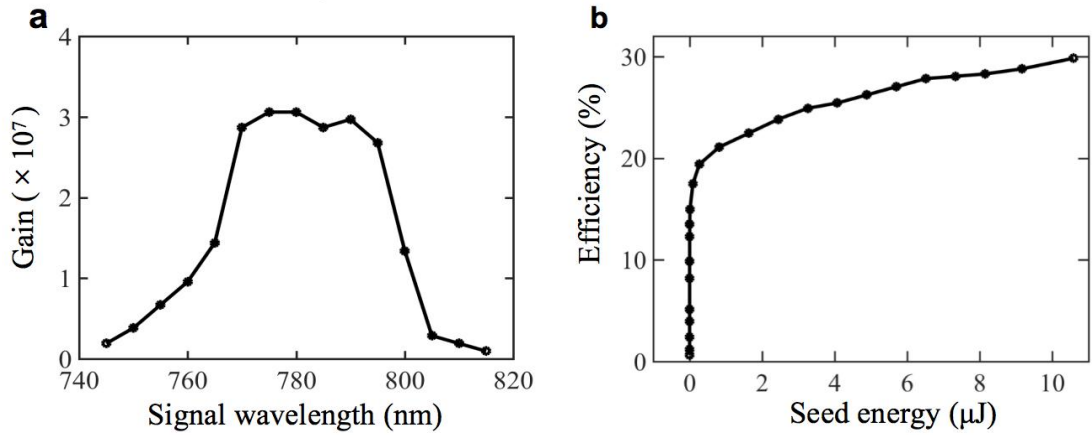

Figure 2 | DUPA gain and efficiency. a, Measured signal gain spectrum under the fixed phase-matching condition in the small-signal regime. b, Measured signal efficiency under the fixed phase-matching condition in the saturated amplification regime.

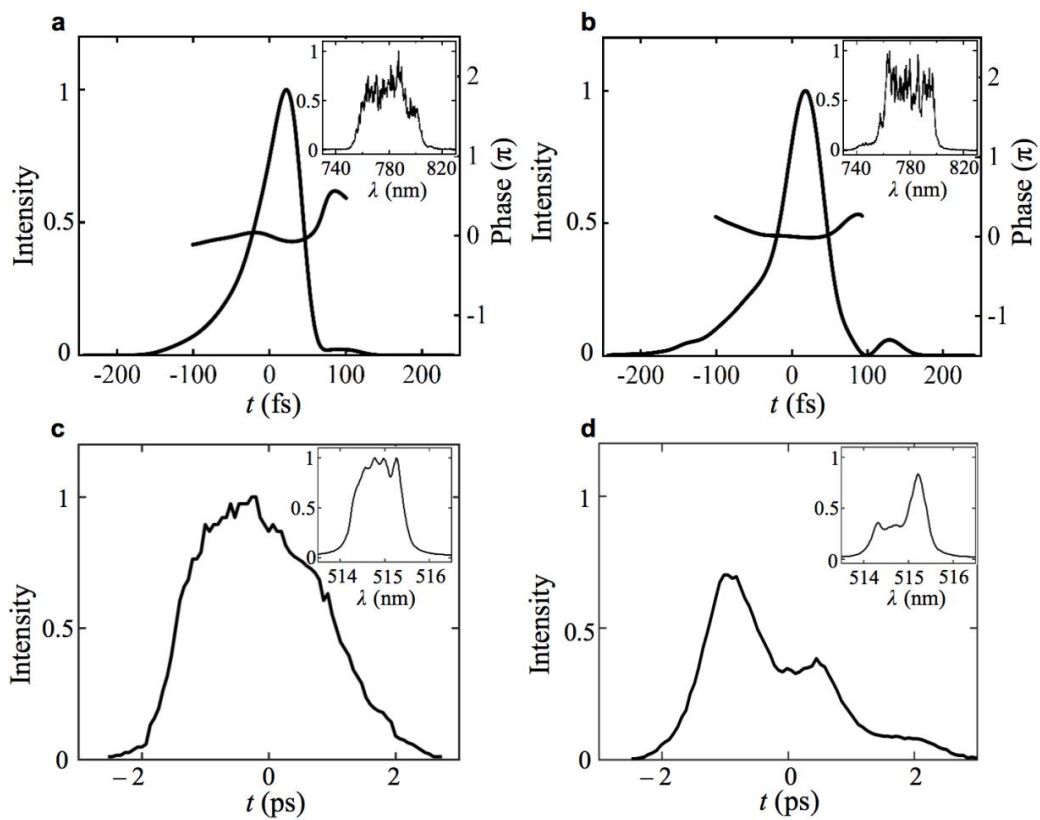

Figure 3 | Amplified signal pulse and depleted pump pulse. a, Measured signal seed pulse and its spectrum at the crystal exit. b, Measured amplified signal pulse and its spectrum at the crystal exit. c, Measured incident pump pulse and its spectrum. d, Measured residual pump pulse and its spectrum at a seed signal energy of 10.5 µJ.

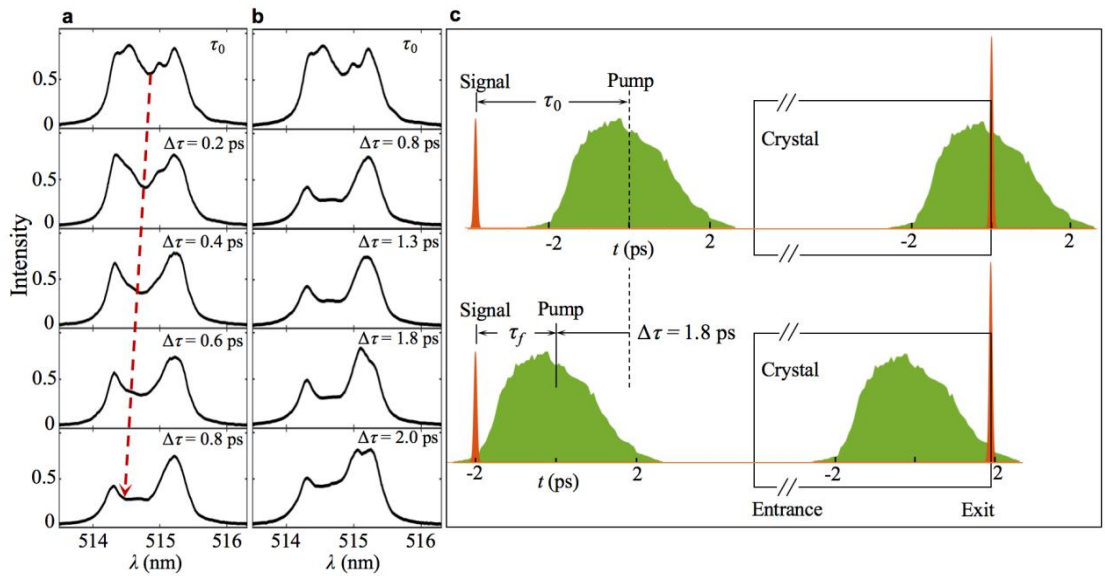

Figure 4 | Characterizations of pump-depletion dynamics and group-velocity mismatch. a and b, Evolution of the residual pump spectrum at a seed signal energy of 3.5 μJ with various delays up to $\Delta\tau$ = 0.8 ps and $\Delta\tau$ = 2.0 ps, respectively. The time delay relative to the first frame is written in each image. The dashed line in the left column shows the spectral shift from the initial to maximal depletion, indicating the pump pulse chirp. c, Timing diagrams for the two specific conditions discussed in the text. The time interval $\tau_0$ before the crystal represents the overall group-velocity mismatch between the pump and signal.